# Density Functional Model for Nondynamic and Strong Correlation

Jing Kong[*] and Emil Proynov

Department of Chemistry and Center for Computational Sciences, Middle Tennessee State University,
1301 Main St., Murfreesboro, TN 37130, USA

**Abstract**

A single-term density functional model for the left-right nondynamic/strong electron correlation is presented based on single-determinant Kohn-Sham density functional theory. It is derived from modeling the adiabatic connection for kinetic correlation energy based on physical arguments, with the correlation potential energy based on the Becke'13 model (Becke, A. D. *J. Chem. Phys*. **2013**, *138*, 074109). This functional satisfies some known scaling relationships for correlation functionals. The fractional spin error is further reduced substantially with a new density-functional correction. Preliminary tests with self-consistent-field implementation show that the model, with only three empirical parameters, recovers the majority of left-right nondynamic/strong correlation upon bond dissociation and performs reasonably well for atomization energies and singlet-triplet energy splittings. This study also demonstrates the feasibility of developing DFT functionals for nondynamic and strong correlation within the single-determinant KS scheme.

## 1. Introduction

Density Functional Theory (DFT) is the major tool in applied theoretical chemistry and material science [1, 2]. Current DFT methods perform quite well when the electronic structure is dominated by a single wavefunction configuration. They are less effective for systems with multiconfigurational character, when nondynamic correlation becomes significant. The term strong correlation is used when the nondynamic correlation becomes dominant. Examples of strongly correlated molecular systems are bond dissociation limits, various transition-metal compounds, diradicals, molecular magnets, to mention just a few. Nondynamic/strong correlation has been and still remains elusive in DFT and its treatment is considered the last frontier [1, 2]. In a stretched bond, approximate DFT correlation potentials are missing a major part of the exact correlation potential that would compensate the multi-center delocalization of the unscreened exact exchange potential. This missing correlation component is called left-right nondynamic correlation [3-6]. The nondynamic correlation has also been analyzed in terms fractional charge and fractional spin [7, 8]. In some cases, the spin-unrestricted approach using a broken symmetry KS single-determinant has been advanced as a practical solution. While often useful, it does not substitute for a fundamentally correct solution, and introduces spin contamination errors.

Various solutions for nondynamic/strong correlation have been proposed, mostly trying to extend KS DFT beyond a single-determinant such that the proper dissociation limit, a degenerate state, can be obtained from the outset. One approach is to utilize multiconfigurations [9-11]. It would dissociate a bond properly, but has the problem of double-counting some of the correlation. Functionals of variable occupation numbers of the KS orbitals have been suggested as a partial remedy [5, 12, 13]. Related to the latter approach, the density matrix functional theory (DMFT) has been pursued [14-16]. Random phase approximation (RPA) [17, 18] has also been used to include some multiconfiguration effects implicitly using Green function and/or response function formalism within KS DFT. Noncolinear DFT methods have been advanced for the description of strongly correlated radicals [19-21] based on time-dependent DFT technique.

On the other hand, it is not obvious how an apparent multiconfiguration state can be accommodated by a single-determinant approach with functional variables based on occupied orbitals. But such an approach has the dual appeal of conceptual simplicity and computational efficiency. After all, the main promise of DFT is its capability of recovering correlation at the cost level of Hartree-Fock (HF). The theoretical justification for this promise is the fundamental theorem of KS-DFT that the exact electron density can be generated by the KS single determinant [22]. Methods of incorporating the nondynamic electron correlation within the single-determinant KS DFT have been proposed in recent years [17, 23-28]. Most notable among them are B05 [23], B13 [24] by Becke, and PSTS by Perdew and co-workers [25]. These methods measure the delocalization of the exact exchange hole in real-space and compensate for this effect with a local correction serving as nondynamic correlation. As shown in refs. [29-31] we have implemented B05 and PSTS self-consistently and efficiently with a resolution-of-identity (RI) scheme and reoptimized the parameters for B05 (named RI-B05 henceforth), with self-consistent field (SCF) demonstrating that they perform similarly to the mainstream methods on standard thermodynamic benchmarks. They perform much better on some strongly correlated molecules, such as the NO dimer and $Cr_2$ [32]. They are based on full HF exchange, contain only a few empirical parameters, and thus rely much less on fortuitous error cancellations. Still, neither method dissociates covalent bonds correctly [31]. Other efforts have been made to correct the fractional-charge and fractional-spin errors that are closely related to strong correlation [33-36]. The most recent advance in tackling strong correlation is the B13 method. This general DFT method is shown to recover the majority of the nondynamic correlation energy, and comes quite close to the dissociation limit. Complementary to these efforts in energy functional modeling, an analytical method has been proposed by us to calculate the population of effectively unpaired electrons for characterizing the nondynamic/strong correlation based on single-determinant representation [32].

---

[*] Correspondence: jing.kong@mtsu.edu.

## 2. Theory

### 2.1 *Basic model for nondynamic correlation*

In this work we propose a new density functional model for left-right nondynamic correlation with strong correlation included. The model starts with the full KS exact exchange, calculated in a HF-like manner. Next we use the adiabatic connection (AC) method to obtain the correlation energy part of the functional, seeking for a form that would properly complement the exact exchange and would take into account nondynamic correlation as well. By virtue of adiabatic connection (AC) [37, 38], the correlation energy is formally expressed as the integral of an AC kernel:

$$E_c = \int_0^1 W_\lambda^c \, d\lambda = \int_0^1 \int w_\lambda^c(r) d^3r \, d\lambda \, , \quad (1)$$

where $\lambda$ ( $0 \leq \lambda \leq 1$) is the AC correlation coupling-strength parameter, $w_\lambda^c$ is the $\lambda$-dependent correlation energy density. The energy kernel $W_\lambda^c$ connects the KS non-interacting reference state and the real correlated state while keeping the electron density the same and independent of $\lambda$. The latter is achieved by using $\lambda$-dependent effective external potential in the KS Hamiltonian. Equation (1) provides, among other things, the inclusion of the correlation kinetic energy given a model for the potential part of the correlation energy functional at full interaction strength, $\lambda = 1$. The total correlation energy in the KS scheme resulting from Eq.(1) is the sum of the correlation kinetic energy and the correlation potential energy at the full interaction strength.

Our starting model of the $\lambda$-dependent energy density for nondynamic correlation has the following Fermi-function-like form:

$$w_\lambda^{nd}(\xi) = a(\xi)\frac{e^{bz(\xi)\lambda} - 1}{e^{bz(\xi)\lambda} + 1}, \quad \xi = (\rho_\sigma, \nabla\rho_\sigma, \nabla^2\rho_\sigma, \tau_\sigma, u_{X\sigma}^{ex}) \quad (2)$$

where $\xi$ is a set of variables in real space, with $\rho_\sigma$ being the electron density for spin σ, $\tau_\sigma$ the KS kinetic energy density, and $u_{X\sigma}^{ex}$ the exact exchange energy density. The factor $a(\xi)$ is a negatively defined real-space function that will be specified below. The coefficient $b$ is a positive empirical parameter that scales the function $z(\xi)$. The latter indicates the relative degree of nondynamic correlation at each point. Electrons have relatively little nondynamic correlation at the point where $z$ is close to zero. The electrons are at the dissociation limit when $z$ approaches infinity. The definition of $z(\xi)$ will be given later.

The design of our model is motivated by some known exact conditions on an AC energy kernel $W_\lambda^c$. First, the energy density $w_\lambda^{nd}$ should decrease monotonically as $\lambda$ increases, since it has been shown that the derivative $dW_\lambda^c/d\lambda$ is always negative [39]. Second, $w_\lambda^{nd}$ should approach a finite value in the limit of $\lambda \to \infty$ because the exact $W_\lambda^c$ has a finite lower bound at that limit [40]. Furthermore, the derivative of $w_\lambda^{nd}$ with respect to $\lambda$ at $\lambda = 0$ should approach minus infinity as $z$ approaches infinity (the dissociation limit). This is due to the exact slope of $W_\lambda^c$ at $\lambda \to 0$ given by the Görling-Levy (GL) perturbation theory as $2E_c^{GL2}$, where $E_c^{GL2}$ is the GL second-order perturbation term [41]:

$$E_c^{GL2} = \sum_{i=1}^{\infty} \frac{\left|\langle \Phi_0 | \hat{V}_{ee} - \hat{v}_1 | \Phi_i \rangle\right|^2}{E_0 - E_i} . \qquad (3)$$

Here $\Phi_i$ are ground ($i = 0$) and excited ($i > 0$) KS determinants, $\hat{V}_{ee}$ is the electron-electron interaction operator and $\hat{v}_1$ is the sum of the classical Coulomb and exact exchange potential operators. The denominators in the $E_c^{GL2}$ expression contain differences between the KS ground state energy $E_0$ and energies from excited-like KS determinants. This makes the slope $2E_c^{GL2}$ approach minus infinity in the dissociation limit where some of these energy differences approach zero due to (quasi) degeneracy that is typical for this limit.

Lastly, $W_\lambda^{nd}$ should be almost a straight line when the correlation is weak, i.e. $z$ is small. Although there is no formal proof of this point, the essence of it has been applied in the development of Becke-Half-and-Half method, the predecessor of B3LYP [42], and has been demonstrated in the AC energy curves for some molecules near equilibrium [43, 44].

The $\lambda$-dependence of $w_\lambda^{nd}(\xi)$ with respect to changing $z$ is qualitatively illustrated in Fig. 1. It resembles the AC energy curves of dissociating hydrogen molecule reported in recent literature (Fig.4 of ref.[43]). The curves go from nearly a straight line at equilibrium (weak correlation) to almost a step function at large inter-nuclear separation (strong correlation).

We determine the yet unspecified function $a(\xi)$ in Eq.( 2) by imposing the following condition at $\lambda = 1$:

$$w_1^{nd}(\xi) = u_c^{nd}(\xi), \qquad (4)$$

where $u_c^{nd}$ is the nondynamic correlation potential energy density. This gives:

$$a = \frac{e^{bz}+1}{e^{bz}-1} u_c^{nd} . \qquad (5)$$

The genesis of $u_c^{nd}$ is the delocalization of the KS exact exchange hole in a multicenter system. When a bond is stretched, the exact exchange hole is distributed between several atoms, resulting in an exchange hole that is too shallow. The nondynamic correlation compensates for this depletion such that the total exchange-correlation (XC) hole becomes localized in a region of about an atomic size. The form of $u_c^{nd}$ used in this work is $(u_{statC}^{opp} + u_{statC}^{par})$ as defined by Eq.(52) in ref.[24]. The main component here is the opposite-spin nondynamic correlation term (see Appendix A):

$$u_{statC}^{opp} = f_{opp}(\mathbf{r}) \left[ \frac{\rho_\alpha(\mathbf{r})}{\rho_\beta(\mathbf{r})} u_{X\beta}^{ex}(\mathbf{r}) + \frac{\rho_\beta(\mathbf{r})}{\rho_\alpha(\mathbf{r})} u_{X\alpha}^{ex}(\mathbf{r}) \right], \qquad (6)$$

where $u_{X\sigma}^{ex}$ is the exact-exchange energy density of spin $\sigma$, $f_{opp}$ is the local correlation factor employed in the B05 and B13 models. It is a complicated, position-dependent function, designed to measure the strength of opposite-spin nondynamic correlation at each point. The details of this formalism are described in the original papers of the B05 [23], and B13 [24] models and our modifications of B05 [30]. A gist of the main formulas relevant to the present model is given in Appendix A.

The real-space function $z(\xi)$ in Eqs.( 2) and ( 5) couples the functional variables with the AC coupling strength parameter $\lambda$. It is a key in the new method, providing a measurement the relative degree of nondynamic correlation at each point in real space. A plausible definition of $z(\xi)$ consistent with the physical motivation behind Eq.( 2) is:

$$z = \frac{u_c^{nd}}{u_c^{dyn}}, \qquad (7)$$

where $u_c^{dyn}$ is the dynamic correlation potential energy density mainly due to the interelectronic cusp condition. The factor $z$ is small when dynamic correlation dominates. When the bond is stretched, nondynamic correlation dominates and $z$ is large. This implicit dependence of a density-dependent variable on the interatomic bond distance is suitable for targeting the features of a $\lambda$-dependent energy density discussed above. This particular choice of $z$ has also some important advantages as explained further on and in Appendix C. The coupling between $z$ and $\lambda$ is further scaled by the parameter $b$, which modulates the transition from a small $z$ to an intermediate one, but has essentially no effect on the dissociation limit when $z$ is large. It is required to be positive and will be determined empirically. The $u_c^{dyn}$ used here is the dynamic correlation potential energy density ($u_{dynC}^{opp} + u_{dynC}^{par}$) from ref.[24] with the modification described in ref.[30]. It corresponds to the potential part of the dynamic correlation at $\lambda = 1$. The dominant component of $u_c^{dyn}$ is the opposite-spin dynamic correlation term (B13):

$$u_{dynC}^{opp}(\rho_\alpha, \rho_\beta) = (1 - f_{opp}) \frac{\rho_\alpha \rho_\beta \Lambda_{opp}^3}{(1 + \Lambda_{opp})}, \qquad (8)$$

where $\Lambda_{opp}$ is dynamic correlation length reflecting the size of the opposite-spin dynamic correlation hole, and $f_{opp}$ is the B05 local correlation factor used in Eq.( 6). The form of the parallel-spin component $u_{dynC}^{par}$ is given in Appendix A.

Integrating the $\lambda$-dependent energy density $w_\lambda^c$ in Eq. ( 1) over $\lambda$ using Eq.( 2) yields the nondynamic correlation energy density. Integrating it over the space and adding the full exact exchange energy $E_x^{ex}$ leads to the following functional:

$$E_{xc} = E_x^{ex} + \int \frac{1 + e^{bz}}{1 - e^{bz}} \left[ 1 - \frac{2}{bz} \ln\left( \frac{1 + e^{bz}}{2} \right) \right] u_c^{nd} dr^3 \equiv E_x^{ex} + \int q_c^{AC} u_c^{nd} d^3r. \qquad (9)$$

We note that $q_c^{AC}$ tends to 0.5 when $z$ approaches 0 (near equilibrium), and tends to 1 when $z$ approaches infinity (the dissociation limit). The correlation kinetic energy $T_c$ can be extracted from Eq.( 9) as:

$$T_c = E_c^{nd} - E_c^{nd}(\lambda=1) = \int q_c^{AC} u_c^{nd} d^3r - \int u_c^{nd} d^3r \; , \qquad (10)$$

where $E_c^{nd}$ is the total nondynamic correlation energy and $E_c^{nd}(\lambda=1)$ is the nondynamic correlation potential energy at the full interaction strength.

It is worth discussing briefly some formal properties of these expressions with respect to known exact conditions. An exact property that the nondynamic correlation energy in Eq. (9) obeys is the finite scaling bound at low-density limit [40]:

$$\lim_{\gamma \to 0} \gamma^{-1} E_c^{nd}[\rho_\gamma] > -\infty \; , \qquad (11)$$

where $\rho_\gamma(\mathbf{r}) = \gamma^3 \rho(\lambda \mathbf{r})$ is the uniformly scaled electron density, $\gamma$ is coordinate scale factor that changes uniformly the length scale of the density while maintaining its normalization. In the context of the AC method, the coordinate scale factor $\gamma$ is the inverse of the coupling strength parameter $\lambda$, $\gamma = 1/\lambda$, and Eq.(11) reflects the strictly correlated limit of $\lambda \to \infty$. The finite scaling bound is a necessary (but not a sufficient) condition for obeying the important Lieb-Oxford bound and other related strict lower bounds on the correlation energy in non-degenerate systems [40, 45]. As a consequence of Eq.(11), the present nondynamic correlation functional scales inhomogeneously with respect to uniform coordinate scaling as shown in Appendix B. These are known exact requirements [46, 47] and fulfilling them by our functional is rather encouraging. We note that the static correlation in B13 does not scale inhomogeneously.

Another exact property of a correlation functional in the case of non-degenerate states is the exact scaling relation at high-density limit [25]:

$$\lim_{\gamma \to \infty} \frac{E_{xc}[\rho_\gamma]}{E_x^{ex}[\rho_\gamma]} = 1 \; , \qquad (12)$$

In the context of AC, Eq.(12) reflects the KS noninteracting limit at $\lambda \to 0$. As shown in Appendix B, our nondynamic correlation functional does not obey this limit. The B05 and B13 functionals also violate this relation as noted in refs.[25, 48].

We have implemented the functional (Eq.(9)) self-consistently and applied it first to the dissociation of $H_2$. The parameter $b$ is 1.2 as explained later, and the basis set is G3Large [49]. The result is shown in Fig. 2 where the new model is named PMF(1) (present model functional with 1 parameter). As one can see, this one-parameter functional provides a major correction to HF, even without dynamic correlation. This is very encouraging because this model is based on a single-determinant, and thus operationally similar to HF. For comparison, we extract a model from the B13 method (post-LDA)[24] that also uses $u_c^{nd}$ only (the static correlation term in B13) and use it to compute the dissociation curve. The curve is noted as B13X in Fig. 2. As one can see, PMF(1) demonstrates a major improvement over this base model.

2.2 *Fractional spin correction and dynamic correlation*

Not shown in Fig.2 is that the PMF(1) dissociation curve will not approach the exact zero limit. This problem is intimately related to the so-called fractional spin error. It dominates at the dissociation limit, and plays a significant role in many molecular systems near equilibrium[8]. In the case of dissociating $H_2$, the electronic structure of each dissociated H atom is two half-electrons with opposite spins, denoted as H $\updownarrow$ . Ideally it must have exactly the same energy as the normal H atom with one electron (denoted as H $\uparrow$). This requires that the local part of the exchange hole in H $\updownarrow$ yield exactly the exchange hole of a half-electron. It then would result in an exchange-correlation hole that is equal to the exchange hole of a whole electron. The method used here for estimating the contribution of the local part of the exact exchange-hole to the potential of the entire hole is to compare the electrostatic potential of a size-relaxed Becke-Roussel exchange-hole (called BR auxiliary hole) with the exact exchange electrostatic potential in real space. Equalization of the two determines the size of BR auxiliary hole, from which the local correlation factor $f_{opp}$ mentioned above, Eqs.( 6) and ( 8), is obtained. Ideally, this size should be exactly half everywhere in H $\updownarrow$. In reality, it is not, but very close. For other atoms, however, the BR auxiliary hole function does not perform as well as for hydrogen, as pointed out in ref.[24].

Solutions of the fractional-spin problem have been proposed in literature. With B13, this problem is treated by altering the curvature of the auxiliary BR hole, with the assistance of an atom-specific parameter, such that $u_{statC}^{opp} + u_{statC}^{par}$ becomes zero for standard atom [24]. In molecular calculations those atomic parameters are partitioned with a real-space weight scheme. Another method has also been proposed recently to correct the fractional spin error for atoms without empirical parameters [34].

Here we propose the following correction to $u_c^{nd}$ in PMF(1):

$$\tilde{u}_c^{nd} = (u_{statC}^{opp} + u_{statC}^{par})\left(1 + \frac{1}{2}\sqrt{\frac{\alpha}{\pi}} \exp(-\alpha/z^2)\left(\left(\frac{D_\alpha}{\rho_\alpha^{5/3}}\right)^{1/3} + \left(\frac{D_\beta}{\rho_\beta^{5/3}}\right)^{1/3}\right)\right), \quad (13)$$

where the parameter $\alpha$ is determined empirically. The function $D_\sigma \equiv \tau_\sigma - \frac{1}{4}\frac{|\nabla\rho_\sigma|^2}{\rho_\sigma}$ is used to eliminate the one-electron self-interaction error [50]. This modification essentially compensates the insufficient recovery of nondynamic correlation towards the dissociation limit. When $z$ is small (negligible nondynamic correlation) the correction tends to zero and the modification has no effect. When $z$ is large, the correction tends to a density-dependent shift. The effect of this correction will be demonstrated below. We note that this correction does not change the scaling property discussed above qualitatively, as explained in Appendix B.

To complete the new model, a dynamic correlation term based on AC is added:

$$E_c = \int q_c^{AC}\, \tilde{u}_c^{nd}\, dr^3 + \int \left(\int_0^1 w_\lambda^{dyn}\, d\lambda\right) dr^3 \ . \quad (14)$$

In this work, the $\lambda$-integrated dynamic correlation energy density is taken from Eqs.(37) and (38) of the B13 paper [24] (see Appendix A). Note that it is different from the dynamic correlation potential energy density, Eq.( 8), used in our definition of the $z$ factor (Eq.( 7)). We also note that the dynamic correlation used in the B13 functional corresponds to $w_{\lambda=1}^{dyn}$. Eq.( 14) contains two empirical parameters ($b$ in $q_c^{AC}$, $\alpha$ in $\tilde{u}_c^{nd}$), and is abbreviated as PMF(2) for present model functional with two parameters.

2.3. *Comparison with B13*

We emphasize that our model takes into account of the correlation kinetic effect based on the explicit AC integration for both nondynamic and dynamic correlations. On the other hand, it shares the same correlation potential energy density with B13. For comparison, we recite the B13 method, perhaps the only general-purpose DFT functional designed for dynamic, nondynamic, and strong correlations. It adds the following correlation energy to the full exact exchange:

$$E_C^{B13} = a_{statC}^{opp}\int u_{statC}^{opp} d^3r + a_{statC}^{par}\int u_{statC}^{par} d^3r + a_{dynC}^{opp}\int u_{dynC}^{opp} d^3r + a_{dynC}^{par}\int u_{dynC}^{par} d^3r$$

$$E_{StrongC}^{B13} = E_C^{B13} + c_2 \int x^2 (u_c^{nd} + u_c^{dyn}) d^3r\,; \qquad x \equiv \frac{u_c^{nd}}{u_c^{nd} + u_c^{dyn}} \ . \quad (15)$$

The sum of the first four terms comprise the static and dynamic correlation as described in the B13 paper. The last term is the correction for strong correlation. The first two terms in B13 are first order with respect to $x$, and the strong correlation correction is second order in a polynomial expansion in $x$. B13 and our model share the potential part of the correlation energy density, but B13 tries to compensate the missing correlation kinetic energy through empirical fitting of five linear parameters. It also employs one parameter for each element for minimization of the fractional spin error. We note that results with and without the strong correlation correction are reported in the B13 paper [24]. In this work we focus on the comparison with B13 with the strong correlation correction included.

There is no rigorous distinction between nondynamic and dynamic correlation in general. However, such a separation may arguably exist in the case of the dissociated $H_2$. A dissociated H atom with fractional spin occupancy( H $\updownarrow$ ) has the same total energy as that of a normal H atom ( H $\uparrow$ ), but only half of the exchange energy of the latter. Therefore the correlation energy is equal to the exchange energy in H $\updownarrow$, i.e. $E_c = E_x^{ex}$. Furthermore, this correlation energy should be purely nondynamic in nature, since the total electron density does not change between the two cases and neither should the electron-electron cusp condition [24]. The $z$ factor defined in Eq.( 7) would then approach infinity in the case of H $\updownarrow$, and the following relationship should be true for the exact correlation energy density $u_c$:

$$\lim_{z \to \infty} u_c = u_x^{ex}. \qquad (16)$$

Our model satisfies this requirement under the ideal condition of no fractional spin error since $q_c^{AC}$ becomes 1 as $z \to \infty$ and $u_c$ becomes $u_x^{ex}$. (See Appendix C for details). Under the same condition, the correlation energy $E_{strongC}^{B13}$ in B13 becomes $1.06 E_x^{ex}$, because $c_2 + a_{staticC}^{opp} = 1.06$ and $x = 1$ in Eq.( 15).

We note that a formal relationship can be drawn between B13 and our model when nondynamic correlation is relatively weak, i.e. $z$ is small. As one can see from Eqs.( 15) and ( 7), $x$ in B13 and $z$ in our model are related as: $z^{-1} = 1 - x^{-1}$. When $z$ is small, $x$ is approximately equal to $z$. A Taylor expansion of the nondynamic correlation term in Eq.( 14) with respect to $z$ up to the second order would then yield a polynomial expansion similar to B13. At the dissociation limit, such a connection does not exist. Under this condition, $x$ is approximately equal to $1 - z^{-1}$ and the requirement specified by Eq.( 16) can be expressed alternatively as $\lim_{(1-x) \to +0} u_c = u_x^{ex}$ in the case of dissociated $H_2$ molecule. A Taylor expansion of $u_c$ with respect to $1 - x$ is impossible since $1 - x$ is discontinuous at $x = 1$ ($x$ cannot be greater than 1 by definition).

## 3. Results and discussion

The preliminary optimization of the parameters of PMF(2) is carried out to minimize the mean fractional spin error and the mean atomization energy error of a set of molecules to maintain a balance between thermochemistry and the dissociation limit. The fractional spin error is calculated as the energy difference between an atom in its normal ground state configuration without spin polarization and in the state with fractional spin occupancy corresponding to its homonuclear dissociation limit. It is a special case of the general fractional spin error [8] and quantitatively measures the error at the dissociation limit. We calculate this error for a few selected first and second row atoms included in the optimization test. The optimized model is then assessed for the atomization energies of a larger set of molecules used in our previous studies [30, 51]. All calculations are done with SCF using G3Large basis and a large unpruned grid.

The results of PMF(2) are listed in Table 1, with the optimized values for parameters $b$ and $\alpha$ in the caption. The results are compared with those of the HF method and the real-space correlation DFT functionals B05, B13 and PSTS that also include full exact exchange. Comparison is also made with commonly used functionals such as nonempirical PBE [52], and parameterized hybrid functionals B3LYP [42] and M06-2X [53]. As one can see, HF gives very large fractional spin errors due to the lack of nondynamic/strong correlation. All other methods besides present models and B13 yield substantial fractional spin errors too, albeit the real-space correlation methods B05 and PSTS perform relatively better than B3LYP and M06-2X. PMF(2) recovers the vast majority of the nondynamic/strong correlation with only two parameters, reducing the error from 234 kcal/mol on average to 8.9 kcal/mol. B13 captures about the same amount of nondynamic/strong correlation with a correction term parameterized for each element. The list of the individual fractional-spin error for each of the atoms included in the test set is given in the Supporting Information.

A general-purpose DFT functional should ideally perform well for both the dissociation and the equilibrium. Table 1 lists the MADs of atomization energies of the present assessment set, calculated with our model and the other DFT methods discussed above. The error of PMF(2) in atomization energy is somewhat larger than with the other DFT methods listed in Table 1, but still represents a tremendous improvement over the HF reference. To further improve the performance for the atomization energy, we introduce another parameter ($c$) that scales the contribution of the parallel-spin component of the nondynamic correlation, i.e. $u_{statC}^{opp} + u_{statC}^{par} \rightarrow u_{statC}^{opp} + cu_{statC}^{par}$ in Eq.( 13). We call this model PMF(3), standing for present model functional with three parameters. All the three parameters are optimized with the same data set as for the optimization of PMF(2), and the performance data of PMF(3) are listed in Table 1. The individual error with PMF(3) for each case in the considered test sets is listed in the Supporting Information. As one can see, the MAD of PMF(3) for the atomization energy is reduced almost by half compared to PMF(2). The fractional spin error with PMF(3) remains about the same as PMF(2). When both errors are measured together, PMF(2) and PMF(3) clearly represent a substantial improvement over the current mainstream DFT functionals. We note that the optimized value of the new parameter $c$ in PMF(3) is close to the ideal value of 1, demonstrating the physical soundness of our method. Performance could be further improved by scaling the opposite and parallel spin components of the dynamic correlation.

To assess further the performance of PMF(2) and PMF(3) for strong correlation, we have calculated the dissociation curves of $H_2$, $F_2$ and $N_2$ and compared them with some other DFT methods, as shown in Figs. 3, 4 and 5. We find that PMF(2), PMF(3) and B13 perform similarly and are much better than HF and other methods such as B3LYP, M06-2X, and the hyper-GGA methods PSTS and MCY [31]. This performance, while still imperfect, is achieved with 100% HF reference and only a few parameters using spin-symmetric single-determinant. The main visible problem for our models is that they approach the dissociation limit from above the zero line through a shallow maximum even with SCF. We are currently working on modifying and developing further the model in order to overcome this problem. We note that shallow maxima on the dissociation curves are not unique to our method and B13. They also appear with other methods that are more computationally complicated, such as RPA approaches with DFT [17, 54] and coupled-cluster with doubles with wavefunction theories [55, 56]. The reason for these maxima seems to be insufficient static correlation in certain intermediate distance range.

One class of chemical system with significant nondynamic correlation is radicals and diradicals. Current DFT functionals are known to often fail in reproducing the singlet-triplet splitting in diradicals even with spin unrestricted DFT formalism. Table 2 lists results for the singlet-triplet splittings of four diradicals well-known for their difficulty. One can see that RHF performs badly, but unrestricted HF or DFT (UTPSS) are also not a cure for this type of problem. PMF(3), on the other hand, perform reasonably well without reoptimization. It is encouraging that PMF(3) delivers essential nondynamic correlation over RHF reference without resorting to unrestricted approach and fortuitous error cancellation. Note that the performance of UTPSS was considerably improved via a combination of scaled TPSS correlation and spin-projected UHF (SUHF) for this type of system [57], the SUHF+$f_c$TPSS column in Table 2.

## 4. Conclusions

In summary, we have presented an adiabatic connection energy density kernel for nondynamic correlation. The derived functional treats nondynamic correlation from equilibrium to the dissociation limit with a single term. A correction is also proposed for the fractional spin error at the bond dissociation limit. The combination of this functional with an AC-coupling-strength integrated dynamic correlation, implemented with SCF, yields a general correlation functional for thermochemistry and strongly correlated systems based on single-determinant representation. The model, with just a few empirical parameters, is able to describe well nondynamic/strong correlation and performs reasonably for atomization energies and singlet-triplet energy splittings without relying on the fortuitous error cancelation with an admixed HF exchange common to many contemporary DFT methods. It also satisfies some known exact relationships for correlation functionals. Besides B13 our study here provides another evidence of the feasibility of developing DFT functionals for nondynamic and strong correlation with conciseness and self-consistency within the single-determinant KS scheme.

**Acknowledgement**: JK is indebted to Dr. Axel Becke and Dr. Preston MacDougall for helpful discussions. The authors thank Dr. Fenglai Liu and Mr. Matthew Wang for discussions and assistance. The work is supported by Middle Tennessee State University.

Supporting information with additional results and details is available free of charge via the Internet at http://pubs.acs.org.

**Appendix A: Summary of the formulas and expressions used in the present models.**

The form of $u_c^{nd} \equiv u_{statC}^{opp} + u_{statC}^{par}$ used in Eqs.(5),(7) in the text is the nondynamic energy density as defined in Eq.(52) of ref.[24]:

$$u_{statC}^{opp} = f_{opp}(\mathbf{r})\left[\frac{\rho_\alpha(\mathbf{r})}{\rho_\beta(\mathbf{r})}u_{X\beta}^{ex}(\mathbf{r}) + \frac{\rho_\beta(\mathbf{r})}{\rho_\alpha(\mathbf{r})}u_{X\alpha}^{ex}(\mathbf{r})\right], \tag{A1}$$

where $u_{X\sigma}^{ex}$ is the exact-exchange energy density of spin $\sigma$, $f_{opp}$ is the B05 local nondynamic correlation factor with the modifications described in ref.[30]:

$$f_{opp}(\mathbf{r}) = \min(f_\alpha(\mathbf{r}), f_\beta(\mathbf{r}), 1), \quad 0 \leq f(\mathbf{r}) \leq 1, \tag{A2}$$

and $N_{X\sigma}^{eff}$ is the B05/B13 relaxed normalization of the exchange hole within a region of roughly atomic size. Theoretically it should not exceed 1, but practically it is allowed to be as large as 2 for numerical and other accuracy related reasons. The same condition is maintained in our models.

$$u_{statC}^{par} = -\frac{1}{2}\sum_\sigma \rho_\sigma(\mathbf{r})A_{\sigma\sigma}(\mathbf{r})M_\sigma^{(1)}(\mathbf{r}), \tag{A3}$$

$A_{\sigma\sigma}$ are second-order same-spin correlation factors, and $M_\sigma^{(1)}$ is the first-order moment of the Becke-Roussel (BR) relaxed exchange hole ($\bar{h}_{X\sigma}^{eff}$) [23] ($n=1$ below):

$$M_\sigma^{(n)}(\mathbf{r}) = 4\pi \int_0^\infty s^{n+2}\left|\bar{h}_{X\sigma}^{eff}(\mathbf{r},s)\right|ds, \tag{A4}$$

$$A_{\sigma\sigma} = \min(A_{1\sigma}, A_{2\sigma}), \quad A_{1\sigma} = \frac{1 - N_{X\sigma}^{eff} - f N_{X(-\sigma)}^{eff}}{M_\sigma^{(2)}}, \quad A_{2\sigma} = \frac{D_\sigma}{3\rho_\sigma}, \quad (\sigma \equiv \alpha, (-\sigma) \equiv \beta \text{ and v.v.}), \tag{A5}$$

$$D_\sigma \equiv \tau_\sigma - \frac{1}{4}\frac{|\nabla\rho_\sigma|^2}{\rho_\sigma}, \quad \tau_\sigma(\mathbf{r}) = \sum_i^{occ}|\nabla\psi_{i\sigma}(\mathbf{r})|^2, \tag{A6}$$

$\tau_\sigma$ is the (gage-free) KS kinetic energy density as defined in [58, 59], $M_\sigma^{(2)}$ is the second-order moment of the Becke-Roussel relaxed exchange hole ($n = 2$ in Eq.(A4)).

The original calculation of the above terms involved solving a non-linear algebraic equation for a certain dimensionless function $x_\sigma(y_\sigma)$ at each point [23]:

$$\frac{(x_\sigma - 2)}{x_\sigma^2}\left(e^{x_\sigma} - 1 - \frac{x_\sigma}{2}\right) = y_\sigma \equiv -\frac{3}{4\pi}\frac{Q_\sigma}{\rho_\sigma^2}U_{X\sigma}^{ex}, \tag{A7}$$

where $(-Q_\sigma)$ is the curvature of the exact exchange hole [59]:

$$Q_\sigma(\mathbf{r}) = \frac{1}{6}\left[\nabla^2\rho_\sigma(\mathbf{r}) - 2D_\sigma(\mathbf{r})\right], \quad D_\sigma = \tau_\sigma - \frac{1}{4}\frac{|\nabla\rho_\sigma|^2}{\rho_\sigma}. \tag{A8}$$

The relaxed exchange-hole normalization and the exchange-hole moments involve this $x(y)$ function:

$$N_{X\sigma}^{\text{eff}} = \left(\frac{2}{3}\right)^{3/2} \pi \rho_\sigma^{5/2} e^{x_\sigma} \left[\frac{(x_\sigma - 2)}{x_\sigma Q_\sigma}\right]^{3/2}, \tag{A9}$$

$$M_\sigma^{(1)} = N_{X\sigma}^{\text{eff}} \left(\frac{\rho_\sigma(x_\sigma - 2)}{6 x_\sigma Q_{s\sigma}}\right)^{1/2} \left[x_\sigma + \frac{4}{x_\sigma} - \left(1 + \frac{4}{x_\sigma}\right) e^{-x_\sigma}\right], \tag{A10}$$

$$M_\sigma^{(2)} = N_{X\sigma}^{\text{eff}} \frac{\rho_\sigma(x_\sigma - 2)}{6 x_\sigma Q_\sigma} (x_\sigma^2 + 12). \tag{A11}$$

Accurate analytic interpolation of $x_\sigma(y_\sigma)$ was presented in ref. [30] and has the form:

(i) In the region ($-\infty \le y \le 0.15$):

$$x(y) = g(y) \frac{P_1(y)}{P_2(y)}, \quad g(y) = 2 \frac{(2 \arctan(a_1 y + a_2) + \pi)}{2 \arctan(a_2) + \pi}, \quad -\infty \le y \le 0.15, \tag{A12}$$

$$P_1(y) = \sum_{i=0}^{5} c_i y^i; \quad P_2(y) = \sum_{i=0}^{5} b_i y^i, \tag{A13}$$

where $a_i$, $b_i$, $c_i$ are interpolation coefficients;

(ii) In region ($-0.1 \le y \le 1001$):

$$x(y) = \frac{R_1(y)}{4 R_2(y)}; \quad R_1(y) = \sum_{i=0}^{6} d_i y^i, \quad R_2(y) = \sum_{i=0}^{6} e_i y^i, \quad 0 \le y \le 1001, \tag{A14}$$

(iii) In region ($1000 \le y \le +\infty$):

$$x(y) = \ln(y)^{\left[1 + \frac{1}{\ln(y)}\right]} + \frac{1.23767}{\ln(y)} + \frac{9.37587}{\ln(y)^2} - \frac{19.4777}{\ln(y)^3} + \frac{13.6816}{\ln(y)^4} - 0.078655. \tag{A15}$$

The original form of the nondynamic correlation factor $f_{opp}$ brings derivative discontinuity. We have suggested to smoothen its form so that the derivatives for the SCF algorithm be feasible [30]:

$$f_\sigma(\mathbf{r}) \leftarrow 1 - \frac{1}{4\delta} (f_\sigma(\mathbf{r}) - 1 - \delta)^2, \quad |f_\sigma(\mathbf{r}) - 1| \le \delta, \ \delta = 0.05, \tag{A16}$$

where values of $f_\sigma(\mathbf{r})$ larger than $f_\sigma(\mathbf{r}) + \delta$ are capped to 1.0.

$$f_{opp} = (f_\alpha - f_\beta) H(p; z) + f_\beta, \tag{A17}$$

$$H(p; z) = \frac{1}{(1 + e^{pz})}, \quad z \equiv \frac{(f_\alpha - f_\beta)}{(f_\alpha^2 + f_\beta^2)}. \tag{A18}$$

$$A_{\sigma\sigma} = t_\sigma H(q; t_\sigma) + A_{2\sigma}, \quad t_\sigma = A_{1\sigma} - A_{2\sigma}, \tag{A19}$$

where $H(q; t_\sigma)$ has the same form as in Eq.(A18) but with arguments, $A_{1\sigma}$ and $A_{2\sigma}$ given by Eq.(A5).

The dynamic correlation potential energy density used in Eq.(7) of the text has the same form as in B13:

$$u_c^{dyn} = u_{dynC}^{opp} + u_{dynC}^{par}, \tag{A20}$$

$$u_{dynC}^{opp}(\rho_\alpha, \rho_\beta) = (1 - f_{opp}) \frac{\rho_\alpha \rho_\beta \Lambda_{opp}^3}{(1 + \Lambda_{opp})}, \tag{A21}$$

where $\Lambda_{opp}$ is dynamic correlation length reflecting the size of the opposite-spin dynamic correlation hole:

$$\Lambda_{opp} = 0.63 \left( \frac{N_{X\alpha}^{eff}(\mathbf{r}) \rho_\alpha(\mathbf{r})}{\left| u_{X\alpha}^{ex}(\rho_\alpha(\mathbf{r})) \right|} + \frac{N_{X\beta}^{eff}(\mathbf{r}) \rho_\alpha(\mathbf{r})}{\left| u_{X\beta}^{ex}(\rho_\alpha(\mathbf{r})) \right|} \right); \tag{A22}$$

$$u_{dynC}^{par} = \sum_\sigma (1 - f_{\sigma\sigma}) \frac{\rho_\sigma D_\sigma \Lambda_{\sigma\sigma}^5}{(1 + \Lambda_{\sigma\sigma}/2)}, \tag{A23}$$

$$\Lambda_{\sigma\sigma} = 1.76 \frac{N_{X\sigma}^{eff}(\mathbf{r}) \rho_\sigma(\mathbf{r})}{\left| u_{X\sigma}^{ex}(\rho_\sigma(\mathbf{r})) \right|}. \tag{A24}$$

In the final energy expression of the present model, Eq.(13) of the text, the AC $\lambda$-integrated dynamic correlation energy density ($\varepsilon_{dynC} = \int_0^1 w_\lambda^{dyn} d\lambda = \varepsilon_{dynC}^{opp} + \varepsilon_{dynC}^{par}$) is different from the one used in B13 and corresponds to Eqs.(37-38b) of ref.[24] instead:

$$\varepsilon_{dynC}^{opp} = -0.8(1 - f_{opp}) \rho_\alpha \rho_\beta \Lambda_{opp}^2 \left[ 1 - \frac{\ln(1 + \Lambda_{opp})}{\Lambda_{opp}} \right], \tag{A25}$$

$$\varepsilon_{dynC}^{par} = -0.03 \sum_\sigma (A_{2\sigma} - A_{\sigma\sigma}) \rho_\sigma^2 \Lambda_{\sigma\sigma}^4 \left[ 1 - \frac{2}{\Lambda_{\sigma\sigma}} \ln(1 + \Lambda_{opp}/2) \right]. \tag{A26}$$

**Appendix B: Behavior of the functional ( 9) with respect to the exact scaling relations ( 12),( 11).**

We first verify whether the nondynamic correlation functional, Eq.(9) of the text (PMF(1)), would obey or not the known exact coordinate-scaling relation at the high density limit:

$$\lim_{\gamma \to \infty} \frac{E_{\text{XC}}[\rho_\gamma]}{E_{\text{X}}^{\text{ex}}[\rho_\gamma]} = 1 , \qquad (B1)$$

with the uniformly scaled electron density:

$$\rho_\gamma(\mathbf{r}) = \gamma^3 \rho(\lambda \mathbf{r}) . \qquad (B2)$$

In the context of the adiabatic connection method (AC), the scale factor $\gamma$ is the inverse of the coupling strength parameter $\lambda$, $\gamma = 1/\lambda$, and Eq.(B1) reflects the Kohn-Sham noninteracting limit of AC at $\lambda \to 0$.

Considering our nondynamic correlation functional on standalone, the exact relation (B1) leads to [25]:

$$\lim_{\gamma \to \infty} \frac{E_c^{nd}[\rho_\gamma]}{E_{\text{X}}^{\text{ex}}[\rho_\gamma]} = 0, \qquad (B3)$$

The scaling of the Kohn-Sham exact exchange energy in the denominator of Eq.(B3) is known:

$$E_{\text{X}}^{\text{ex}}[\rho_\gamma] = \gamma E_{\text{X}}^{\text{ex}}[\rho] \equiv \gamma \int u_{\text{X}}^{\text{ex}}(\mathbf{r}) d\mathbf{r} , \qquad (B4)$$

where $u_{\text{X}}^{\text{ex}}(\mathbf{r})$ is the exact-exchange energy density $u_{\text{X}}^{\text{ex}}(\mathbf{r}) = u_{\text{X}\alpha}^{\text{ex}}(\mathbf{r}) + u_{\text{X}\beta}^{\text{ex}}(\mathbf{r})$. It follows from Eq.(B4) that

$$u_{\text{X}}^{\text{ex}}[\rho_\gamma(\mathbf{r})] = \gamma^4 u_{\text{X}}^{\text{ex}}[\rho(\gamma \mathbf{r})] . \qquad (B5)$$

Considering next the numerator of Eq.(B3), we write the nondynamic correlation functional ( 9) as (for closed-shell case, the scaling properties in the general case should be the same):

$$E_c^{nd} = \int q_c^{AC}(\rho_\alpha) f_{opp} \, 2 u_{\text{X}\alpha}^{\text{ex}}(\rho_\alpha) d^3 r , \qquad \rho_\alpha = \rho_\beta = \frac{\rho}{2} . \qquad (B6)$$

Here $f_{opp}$ is Becke's nondynamic correlation factor described in Appendix A and the AC prefactor $q_c^{AC}$ in Eq.(B6) reads (from Eq.(9) in the text):

$$q_c^{AC} = \left(\frac{1+e^{bz}}{1-e^{bz}}\right)\left[1 - 2\frac{\ln(\frac{1+e^{bz}}{2})}{bz}\right], \qquad (B7)$$

where $z$ is the local factor of nondynamic correlation used in Eq.(7) of the text. In the case of closed shell this factor takes the form :

$$z = \frac{2 f_{opp} u_{\text{X}\alpha}^{\text{ex}}(\rho_\alpha)}{u_{dynC}^{opp}(\rho_\alpha, \rho_\beta)} , \qquad (B8)$$

$u_{dynC}^{opp}$ is opposite-spin dynamic correlation potential energy density, (Eq.(A21)).

The opposite-spin dynamic correlation length $\Lambda_{opp}$ in Eq.(A21) in the closed-shell case is:

$$\Lambda_{opp} = 1.26 \frac{N_\alpha^{eff}(\mathbf{r})\rho_\alpha(\mathbf{r})}{|u_{X\alpha}^{ex}(\rho_\alpha(\mathbf{r}))|} \quad . \tag{B9}$$

To obtain the coordinate-scaled forms required for verifying the scaling relation (B3) it suffice to know the scaling of the $z$ factor, Eq.(B8). The scaled form of the numerator in Eq.(B8) follows from Eq.(B5) and the fact that Becke's correlation factor $f_{opp}$ is scale invariant. To obtain the scaled form of the denominator in Eq.(B8), it suffice to find how the dynamic correlation length $\Lambda_{opp}$ scales. From Eqs.(B9), (B2), and (B5) we have:

$$\Lambda_{opp}[\rho_\gamma(\mathbf{r})] = \frac{1}{\gamma}\Lambda_{opp}[\rho(\gamma\mathbf{r})] \quad , \tag{B10}$$

and the scaled form of Eq.(A21) becomes ($\rho_\alpha = \rho_\beta = \rho/2$):

$$u_{dynC}^{opp}(\rho_{\gamma\alpha}(\mathbf{r}),\rho_{\gamma\beta}(\mathbf{r})) = (1-f_{opp})\gamma^3 \frac{\rho_\alpha(\gamma\mathbf{r})\rho_\beta(\gamma\mathbf{r})\Lambda_{opp}^3[\rho_\alpha(\gamma\mathbf{r})]}{\left(1+\frac{1}{\gamma}\Lambda_{opp}[\rho_\alpha(\gamma\mathbf{r})]\right)} \quad . \tag{B11}$$

Using Eqs.(B5), (B10), (B11) in Eq.(B8), the scaled form of our correlation factor $z$ is:

$$z[\rho_{\alpha\gamma}(\mathbf{r}),\rho_{\beta\gamma}(\mathbf{r})] = 2\frac{f_{opp}u_{X\alpha}^{ex}(\rho_\alpha(\gamma\mathbf{r}))}{(1-f_{opp})} \cdot \frac{(\gamma+\Lambda_{opp}[\rho_\alpha(\gamma\mathbf{r})])}{\rho_\alpha(\gamma\mathbf{r})\rho_\beta(\gamma\mathbf{r})\Lambda_{opp}^3[\rho_\alpha(\gamma\mathbf{r})]} \equiv z^{(\gamma)} . \tag{B12}$$

It then follows:

$$\lim_{\gamma\to\infty} z[\rho_{\alpha\gamma}(\mathbf{r}),\rho_{\beta\gamma}(\mathbf{r})] \to \infty \quad , \quad f_{opp} \neq 0 \quad . \tag{B13}$$

The scaled total nondynamic correlation energy in Eq.(B3) then becomes (using Eq.(B5):

$$E_c^{nd}[\rho_{\gamma\alpha}] = 2\gamma\int f_{opp}(\frac{1+e^{bz^{(\gamma)}}}{1-e^{bz^{(\gamma)}}})[1-2\frac{\ln(\frac{1+e^{bz^{(\gamma)}}}{2})}{bz^{(\gamma)}}]u_{X\alpha}^{ex}(\rho_\alpha(\mathbf{r}))d\mathbf{r} \quad . \tag{B14}$$

Equations (B12), (B14) show that this correlation functional generally scales inhomogeneously (non-polynomially) with respect to uniform coordinate scaling with finite values of $\gamma$ as it should be [46]. Using Eqs.(B14) and (B4) in the scaling relation (B3) leads to :

$$\lim_{\gamma\to\infty}\frac{E_c^{nd}[\rho_\gamma]}{E_X^{ex}[\rho_\gamma]} = \lim_{\gamma\to\infty,z^{(\gamma)}\to\infty} 2\int\frac{f_{opp}(\frac{1+e^{bz^{(\gamma)}}}{1-e^{bz^{(\gamma)}}})[1-2\frac{\ln(\frac{1+e^{bz^{(\gamma)}}}{2})}{bz^{(\gamma)}}]u_{X_\alpha}^{ex}(\rho_\alpha(\mathbf{r}))}{E_X^{ex}[\rho_\alpha(\mathbf{r})]}d\mathbf{r} \quad , \tag{B15}$$

$$= 2\int\frac{f_{opp}(-1)[1-2]u_{X_\alpha}^{ex}}{E_X^{ex}}d\mathbf{r} = \frac{1}{E_X^{ex}}\int 2f_{opp}u_{X_\alpha}^{ex}d\mathbf{r} \equiv \frac{1}{E_X^{ex}}\int f_{opp}u_X^{ex}d\mathbf{r}.$$

With the present form of the nondynamic correlation factor, $f_{opp}$, Eq.(A2), our nondynamic correlation functional (9) does not obey in general the exact scaling relation of high-density limit, Eq.(B3), except when the integral on the r.h.s. of Eq.(B15) would vanish occasionally.

Another exact relation that is a stringent test of approximate functionals is the generalized Lieb-Oxford bound by Levy and Perdew [40]:

$$\lim_{\gamma \to 0} \gamma^{-1} E_{XC}[\rho_\gamma] \geq -C \int \rho^{4/3}(\mathbf{r}) d\mathbf{r} \quad . \tag{B16}$$

This property is in the low density limit, which is equivalent to the strictly correlated limit of adiabatic connection, $\lambda = \gamma^{-1} \to \infty$. A step toward verifying the above relation in our case of exchange-plus-correlation is to show that our scaled nondynamic correlation functional multiplied by $\gamma^{-1}$ remains finite on stand alone in this limit.

$$\lim_{\gamma \to 0} \gamma^{-1} E_C^{ND}[\rho_\gamma] > -\infty \quad . \tag{B17}$$

From Eq.(B12) we have:

$$\lim_{\gamma \to 0} z[\rho_{\alpha\gamma}, \rho_{\beta\gamma}] = \frac{2}{(1-f_{opp})} \cdot \frac{f_{opp} u_{X\alpha}^{ex}}{\rho_\alpha \rho_\beta \Lambda_{opp}^2} , \tag{B18}$$

which has a finite value for $f_{opp} \neq 1$. As long as the factor $z^\gamma$ remains finite, our scaled nondynamic correlation energy multiplied by $\gamma^{-1}$ remains finite in the low-density limit, with a finite lower bound as follows from Eqs.(B14),(B17). We show below that it remains finite also in the two special cases:

(i)   $f_{opp} = 0$:

In this case $z \equiv 0$ and taking the limit on the l.h.s. in Eq.(B17) leads to the situation:

$$\lim_{\gamma \to 0} \gamma^{-1} E_c^{ND}[\rho_{\gamma\alpha}] = 2\int \frac{0}{0}[1 - 2\frac{0}{0}]u_{X_\alpha}^{ex}(\rho_\alpha(\mathbf{r}))d\mathbf{r}, \quad f_{opp} = 0. \tag{B19}$$

The second 0/0 case on the r.h.s of the above equation can be resolved using the L'Hopital's rule:

$$\lim_{z^{(\gamma)} \to 0} 2\frac{\ln(\frac{1+e^{bz^{(\gamma)}}}{2})}{bz^{(\gamma)}} = \lim_{z \to 0} 2\frac{\left[\ln(\frac{1+e^{bz^{(\gamma)}}}{2})\right]'}{[bz^{(\gamma)}]'} = \frac{1}{2} \quad , \tag{B20}$$

which leads to:

$$\lim_{\gamma \to 0} \gamma^{-1} E_c^{ND}[\rho_{\gamma\alpha}] = 2\int \frac{0}{0}[1-1]u_{X_\alpha}^{ex}(\rho_\alpha(\mathbf{r}))d\mathbf{r} = 0, \quad f_{opp} = 0. \tag{B21}$$

(ii)   $f_{opp} = 1$:

In this case the factor $z = \infty$ and the result on the r.h.s. of Eq.(B17) is equal to zero for the same reason as for Eq.(B15). Since the exact exchange alone obeys the Lieb-Oxford bound for finite systems, having the limit in Eq.(B17) equal to zero in the latter two particular cases does not contradict to the exact general relation (B16). In the case of $0 < f_{opp} < 1$, the degree to which our functional obeys or not the generalized bound of Levy-Perdew can be only verified numerically, due to the complexity of our functional form.

Let us finally note that adding the fractional spin correction as in Eq.( 13), does not change qualitatively the scaling properties of the resulting functional PMF(2) compared to PMF(1). That is because the density dependent term in this correction is dimensionless and does not change upon coordinate scaling, while the scaling of the exponential term $\frac{1}{2}\sqrt{\frac{\alpha}{\pi}}\exp(-\alpha/z^2)$ is solely due to the scaling of the $z$ factor, Eq.(B12). As it was shown above, this factor goes to infinity when the scale parameter $\gamma$ goes to infinity, then the effect of the coordinate scaling of the fractional spin correction is just a constant energy shift with respect to the PMF(1) scaled form. In the opposite limit of $\gamma$ going to zero, $z$ remains a finite-valued function and the effect of the scaling of the fractional spin correction is just changing the value of the lower bound in Eq.(B17), while it still remains finite.

**Appendix C: Proof that our model satisfies Eq.(16) of the text under an ideal condition.**

We use ($u_{statC}^{opp} + u_{statC}^{par}$), defined in Eq.(52) of ref.[24], as the potential part of the nondynamic correlation ($u_c^{nd}$ in Eq.(4) of the paper). In the case of H$_2$ dissociation, $u_c^{nd}$ is equivalent to $u_{statC}^{opp}$, since $u_{statC}^{par}$ is zero for closed-shell systems. The expression for the computation of $u_{statC}^{opp}$ for a closed-shell system is:

$$u_{statC}^{opp} = \frac{(1 - N_x^{eff})}{N_x^{eff}} u_x^{ex}. \tag{C1}$$

In the above equation, $u_x^{ex}$ is the spin-summed exact exchange energy density as in Eq.(10) of the paper. Note that in ref.[24] in the paper the exact exchange energy density for each spin direction is expressed as the electron density times the Slater potential of the same spin. $N_x^{eff}$ is the effective normalization of the local part of the exact exchange hole of either spin. At the dissociation limit of H$_2$, the ideal value of $N_x^{eff}$ for each dissociated H atom (H$\updownarrow$) should be 0.5, i.e. zero fractional spin error, which leads to:

$$u_c^{nd} = u_x^{ex}. \tag{C2}$$

For our model functional under the same ideal condition, the correction for the fractional-spin error is unnecessary. PMF(2) and PMF(3) are then reduced to PMF(1), and the corresponding correlation energy density becomes the integrand in Eq.(6) of the paper:

$$u_c = q_c^{AC} u_c^{nd}. \tag{C3}$$

For an H$\updownarrow$, the dynamic correlation is zero, and thus $z$ approaches infinity. Due to the fact that $\lim_{z \to \infty} q_c^{AC} = 1$, $u_c$ in our model then becomes:

$$u_c = u_x^{ex}. \tag{C4}$$

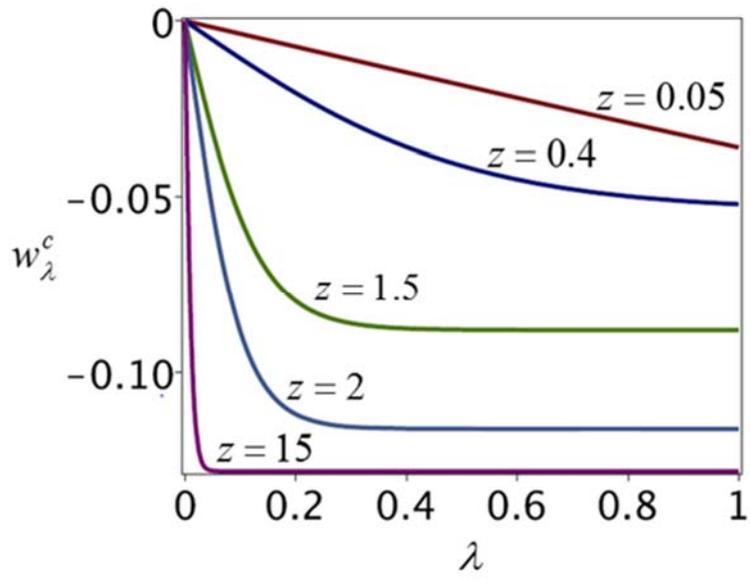

**Figure 1**. Simulated AC curves at various strength of $z$ based on Eq.( 2)

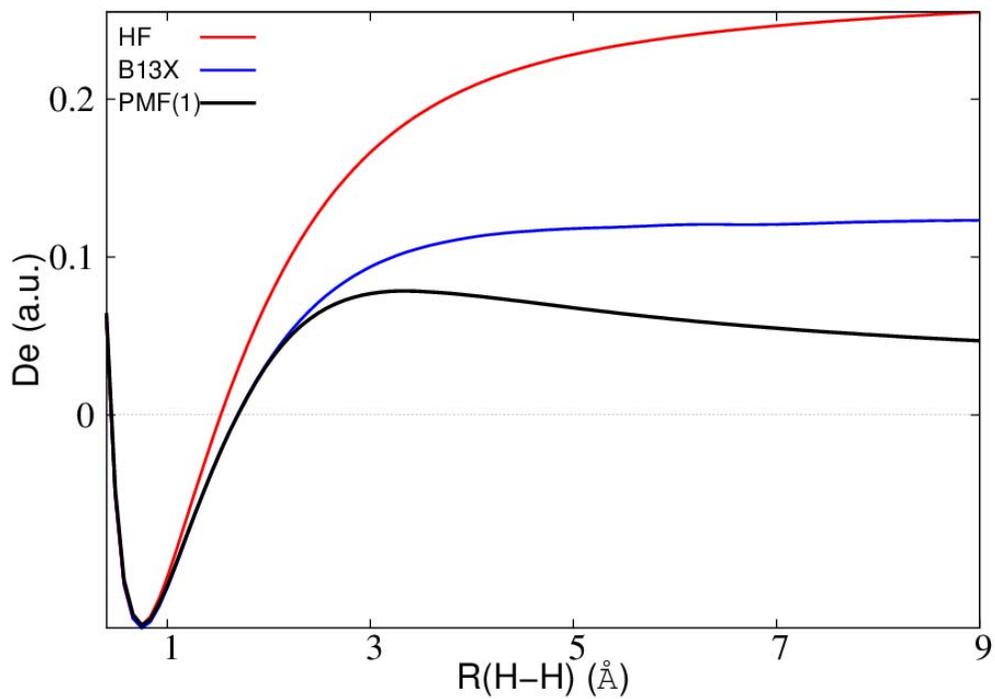

**Figure 2**. Dissociation of $H_2$ with PMF(1), HF and B13X. $D_e$ is the energy difference with respect to twice the energy of the isolated H atom.

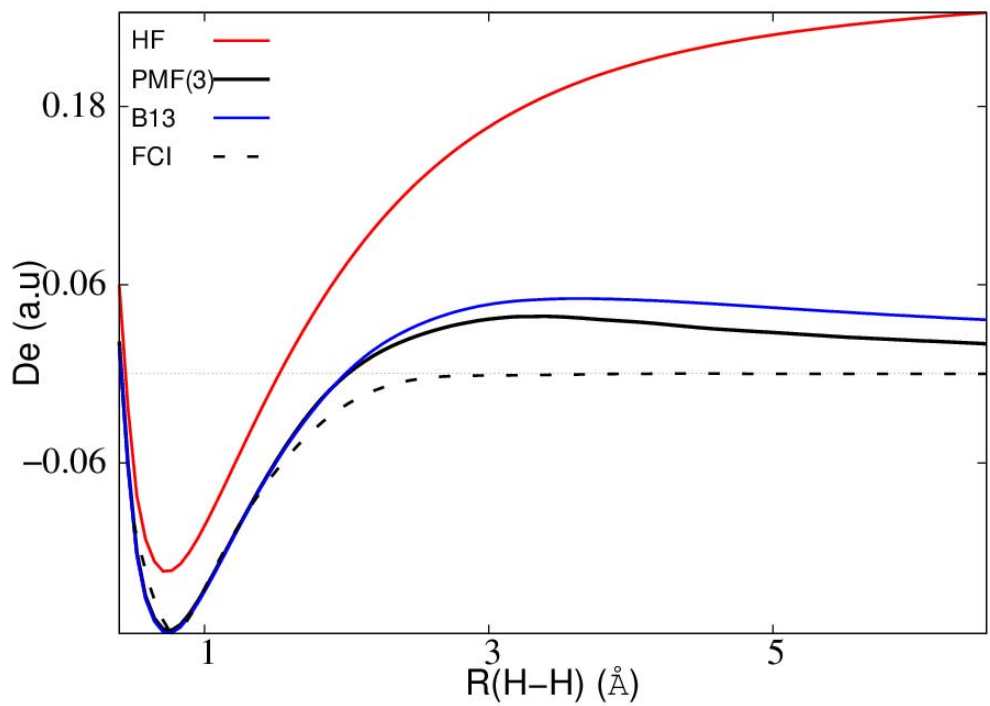

**Figure 3**. Comparison of dissociation curves of $H_2$ among B13, PMF(3), HF and full-configuration-interaction (FCI). $D_e$ is relative to the sum of the energies of two isolated H atoms.

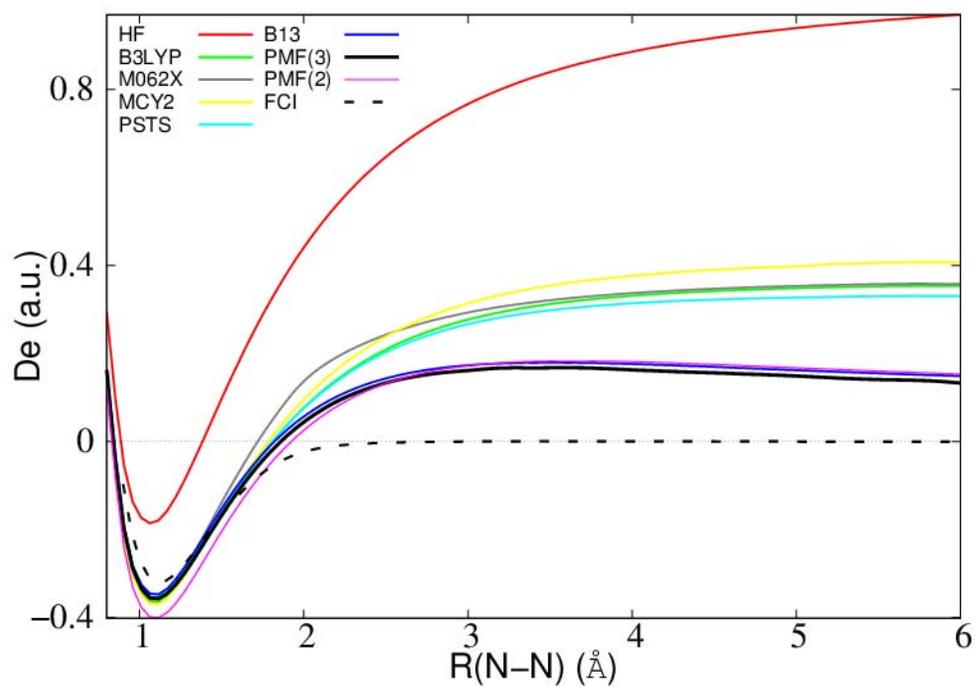

**Figure 4**. Comparison of dissociation curves of $N_2$ among various methods. $D_e$ is relative to the sum of the energies of two isolated N atoms. The full CI (FCI) result is from data available in ref. [60] calculated with 6-31G* basis.

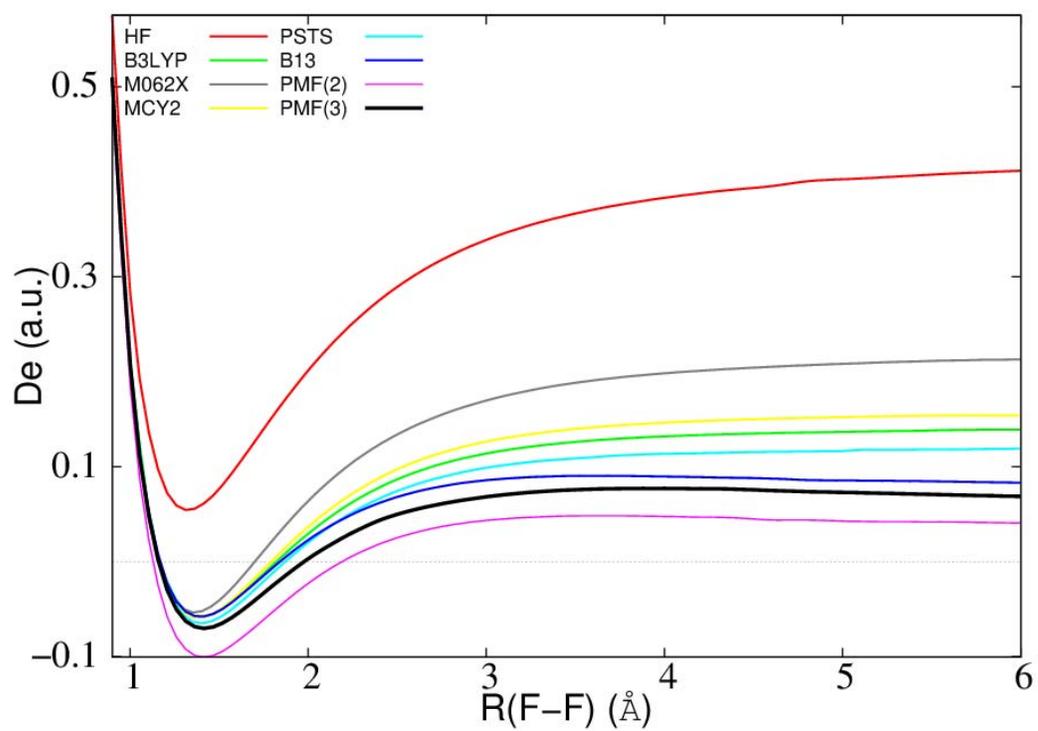

**Figure 5**. Comparison of dissociation curves of $F_2$ among various methods. $D_e$ is relative to the sum of the energies of two isolated F atoms.

**Table 1**. Mean absolute deviations (MADs) of fraction-spin error and atomization energy (kcal/mol) for the assessment set. Individual errors are provided in the Supporting Information. PMF(2) parameters: $b = 1.2, \alpha = 0.037$. PMF(3) parameters: $b = 1.355, \alpha = 0.038, c = 1.128$

| MAD | HF | PMF(2) | PMF(3) | RI-B05 | B13 | PSTS | PBE | B3LYP | M06-2X |
|---|---|---|---|---|---|---|---|---|---|
| Fractional-spin error | 179.21 | 8.94 | 8.14 | 55.95 | 8.78 | 50.97 | 81.50 | 57.56 | 92.17 |
| Atomization energy | 107.37 | 6.30 | 3.64 | 2.32 | ~3.90* | 5.18 | 12.74 | 2.99 | 1.98 |
| Average | 143.29 | 7.62 | 5.89 | 29.14 | ~6.34 | 28.08 | 47.12 | 30.28 | 47.08 |

\* For heteronuclear molecules only $E_C^{B13}$ of B13 method is implemented due to the complexity of the element-wise parameterization for the $E_{StrongC}^{B13}$ term. The MAD with $E_C^{B13}$ on this set is 3.41 kcal/mol, from which we estimate the MAD for B13 with $E_{StrongC}^{B13}$ included using the difference in MAD between these two versions of B13 reported in ref.[24] for the G2 set.

**Table 2**. Singlet-triplet energy splittings (kcal/mol).

| System | RHF[a] | UHF[b] | PMF(3) | UTPSS[b] | SUHF+$f_c$TPSS[b] | Exact[b] |
|---|---|---|---|---|---|---|
| $CH_2$ | 28.2 | 16.9 | 14.5 | 10.3 | 10.0 | 9.4 |
| $NH_2^+$ | 50.4 | 30.5 | 32.5 | 23.6 | 30.2 | 30.0 |
| $PH_2^+$ | -0.6 | -3.7 | -23.6 | -11.1 | -19.0 | -17.3 |
| $SiH_2$ | -4.1 | -6.9 | -23.9 | -15.4 | -20.5 | -21.2 |
| MAD | 18.3 | 9.0 | 4.1 | 4.8 | 0.8 | |

[a] RHF for singlet state, UHF for triplet state.
[b] ref.[57].

For Table of Contents Only

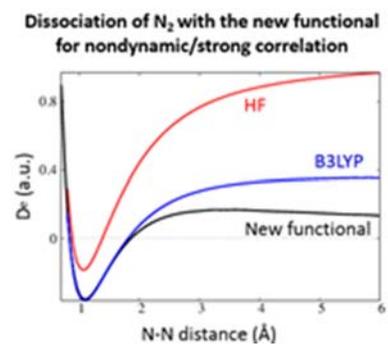